\documentclass[pra,floatfix,amsmath,groupedaddress, twocolumn]{revtex4}
\usepackage{amssymb}
\usepackage{graphicx}
\usepackage{graphics}
\usepackage{amsmath}
\usepackage{mathtools}
\usepackage{amsthm}
\usepackage{color}

\newcommand{\bea}{\begin{eqnarray}}
\newcommand{\eea}{\end{eqnarray}}

\def\C{\hbox{$\mit I$\kern-.7em$\mit C$}}
\def\R{\hbox{$\mit I$\kern-.6em$\mit R$}}
\def\N{\hbox{$\mit I$\kern-.6em$\mit N$}}

\def\ket#1{|#1\rangle}
\newcommand{\one}{\mbox{$1 \hspace{-1.0mm}  {\bf l}$}}
\def\tr{\mathrm{tr}}
\def\ket#1{\left| #1\right>}
\def\bra#1{\left< #1\right|}

\newcommand{\proj}[1]{\ket{#1}\bra{#1}}

\DeclareMathOperator{\BB}{\mathcal{B}}

\newtheorem{theorem}{Theorem}

\newtheorem{observation}[theorem]{Observation}

\begin{document}

\title{Certifying the purity of quantum states with temporal correlations} 

\author{Cornelia Spee}
\affiliation{Institute for Quantum Optics and Quantum Information (IQOQI),
Austrian Academy of Sciences, Boltzmanngasse 3, 1090 Vienna, Austria}
\affiliation{Naturwissenschaftlich-Technische Fakult\"at, 
Universit\"at Siegen, Walter-Flex-Stra{\ss}e 3, 57068 Siegen, Germany}

\date{\today}             

\begin{abstract}
Correlations obtained from sequences of measurements have been employed to distinguish among different physical theories or to witness the dimension of a system. In this work we show that they can also be used to establish semi-device independent lower bounds on the purity of the initial quantum state or even on one of the post-measurement states. For single systems this provides  information on the quality of the preparation procedures of pure states or the implementation of measurements with anticipated pure post-measurement states. For joint systems one can combine our bound with results from entanglement theory to infer an upper bound on the concurrence based on the temporal correlations observed on a subsystem. \end{abstract}

\pacs{03.65.Ta}
\maketitle

\section{Introduction}
Many applications in quantum information theory such as teleportation \cite{teleport} and measurement-based quantum computation \cite{mbqc} use as a  resource pure entangled states. That is, ideally the corresponding protocols are applied to the respective pure resource state and deviations from this resource may result in errors \cite{teleport,faulttolmbqc} and lead to the need of entanglement purification \cite{distillME,distillg,distillg2} or fault tolerant implementations (see e.g. \cite{faulttolmbqc,faulttolmbqc2,faulttolmbqc3}) if one takes into account also imperfections after the preparation procedure. Due to interactions with the environment in experiments often mixed states are prepared instead of the desired pure state. By knowing how much the prepared state differs from a pure state one obtains some intuition on the quality of the preparation process without using full tomography. However, it should be noted that the purity only provides information about how much the prepared state deviates from a pure one (which might not necessarily be the desired one). 

The purity of a quantum state can be quantified via \bea
\mathcal{P}(\varrho)=\tr [(\varrho)^2].
\eea
The purity attains its maximal value of $1$ for pure states and its minimal value of $1/d$ for the maximally mixed state for $d$-dimensional systems. It is related to the linear entropy $S_L(\varrho)= 1-\mathcal{P}(\varrho)$ and the Renyi-2 entropy \cite{renyi} $\mathcal{H}_2(\varrho)=-\log_2 (\mathcal{P}(\varrho))$. The purity (or non-uniformity) of quantum states has been also studied from a resource-theoretic point of view \cite{resourceth,renyi2,resoureth}. Moreover, the task of distilling local pure states via a  subclass of local operations and classical communication has been considered \cite{distillpure, distillpure2}.

It is well known that the purity (of subsystems) of bipartite systems and their entanglement are connected. States for which the purity of the whole system is sufficiently small have to be separable, as there exists a set containing only separable states around the maximally mixed state which has a finite volume \cite{finitevolume1,finitevolume2}.  For two-qubit pure states any entanglement measure can be written as a function of the purity of one of its subsystems as in this case the purity uniquely determines the set of Schmidt coefficients and any entanglement measure for bipartite pure states is a function of the Schmidt coefficients \cite{enmeSC}. The optimal  strategy to estimate the entanglement of an unknown two-qubit pure state from $n$ copies of this state has been shown to correspond to the estimation of  the purity of the single-qubit reduced state  and an explicit optimal protocol  to do so has been proposed \cite{optimal1}. Therefore, this scheme only requires local measurements of one of the parties (but which act non-locally on the different copies). Moreover, in the asymptotic regime separable measurements  of one of the parties assisted by classical communication among the copies can be shown to perform optimally \cite{optimal2}.

For mixed (or higher-dimensional) states the relation among entanglement and sub-system purity is no one-to-one correspondence anymore, however, for example lower \cite{Concboundl} and upper \cite{Concbound} bounds based on the purity of a subsystem and total system have been shown for the concurrence $C(\varrho)$ \cite{conc, conc2}, which is an entanglement measure. 
In particular, it has been shown that \cite{Concboundl}
\bea\label{eqlbound}
\max_{X\in\{A, B\}}2\{\tr [(\varrho)^2]-\tr [(\varrho_X)^2]\}\leq [C(\varrho)]^2\eea
 and \cite{Concbound}
 \bea \label{equbound}
 [C(\varrho)]^2\leq \min_{X\in\{A, B\}}2\{1-\tr [(\varrho_X)^2]\},
 \eea
where $\varrho_X$ is the reduced state of subsystem $X$. The first bound captures quantitatively the observation that only for entangled states the reduced states can be more mixed than the state of the whole system \cite{horodeck}. The upper and lower bound on the concurrence given above can be determined in an experiment by measuring local observables using two identical copies of the state $\varrho$ \cite{Concboundl,Concbound}. The purity (or Renyi-$n$ entropies) of a system can also be experimentally measured by employing two copies of the state (see e.g. \cite{exppur,exppur3,exppur5,exppur2,exppur4} and references therein) or by performing randomized measurements \cite{randmeas1,randmeas2,randmeas3,randmeas4}. It has been shown that if one uses two copies, a non-local unitary among them and a local two-outcome measurement on only one of the copies  but no ancilla or randomized measurements it is only possible to extract the purity  in case the dimension is odd \cite{odddim}. The task of discriminating pure and mixed states has been considered \cite{discrimination1, discrimination2}  which also lead to schemes to estimate the purity. These are either based on maximum confidence discrimination \cite{discrimination1} or an uncertainty relation \cite{discrimination2} and require non-local measurements among the copies or control over the measured observable. Moreover, measurement schemes that allow to determine the purity of single-mode Gaussian states have been proposed \cite{Gaussian} and the relation among the (global and local) purities and entanglement of Gaussian states has been studied \cite{Gaussian2}. 

By performing tomography on the system one could reconstruct the state and calculate the purity of the system.  In particular, there exist adaptive schemes which do not rely on any assumption on the states \cite{adaptive, adaptive2} or which are designed for pure states and in which the assumption of purity can be certified from the observed data \cite{adaptivepure1, adaptivepure2}. However, it should be noted that as in any tomographic approach the measurements are required to be characterized (at least to some extent). The relation of the scaling of the accuracy in device-dependent adaptive process tomography and the purity of the measured state has been studied \cite{adaptive3}.

Device-independent bounds on the linear entropy (of the total system) or the concurrence can be also obtained from the value of violation of a Bell inequality \cite{boundvioBI,boundvioBI2}. Moreover, device-independent entropy witnesses based on dimension witnesses have been proposed in the context of prepare-and-measure scenarios \cite{witboundMP} and sector lengths which are related to the average purity of reduced states have been studied (see e.g. \cite{seclength1, seclength2, seclength3} and references therein). 

Here we propose to use the temporal correlations obtained from sequences of measurements on a single copy to deduce a semi-device-independent lower bound on the purity. This approach relies only on the assumption of the dimension of the measured (sub)system and that measurements can be repeated (see below for more details). Note that even though for a single qubit system less measurements are required in a tomographic approach  than in our approach such schemes require knowledge about the measurements that are implemented. Moreover, our approach does not require to prepare two identical copies of the state at the same time and to act non-locally on the subsystems of different copies \footnote{The observables that one needs to measure in order to obtain the bounds in Eqs. (\ref{eqlbound}) and (\ref{equbound}) are local with respect to the splitting $A_1A_2|B_1B_2$ where $A_i$ ($B_i$) refers to party $A$ ($B$) of copy $i$ respectively but are non-local with respect to the splitting $A_1|A_2$ and/or $B_1|B_2$ \cite{Concboundl,Concbound}.}. It is straightforward to see from the equations above that a lower bound on the purity of a (sub)system provides an upper bound on the linear entropy or the concurrence. Moreover, it has been shown that a lower bound on the purity implies a lower bound on the accessible information \cite{accessinfo}.

Our approach uses sequential measurements and is conceptually different from the ones previously studied. In particular, we can also give a lower bound on the maximal purity of the post-measurement state at the second time step for one of the outcomes provided the purity of the initial state is known. 

\begin{figure}
\begin{center}
\includegraphics[width=1\columnwidth]{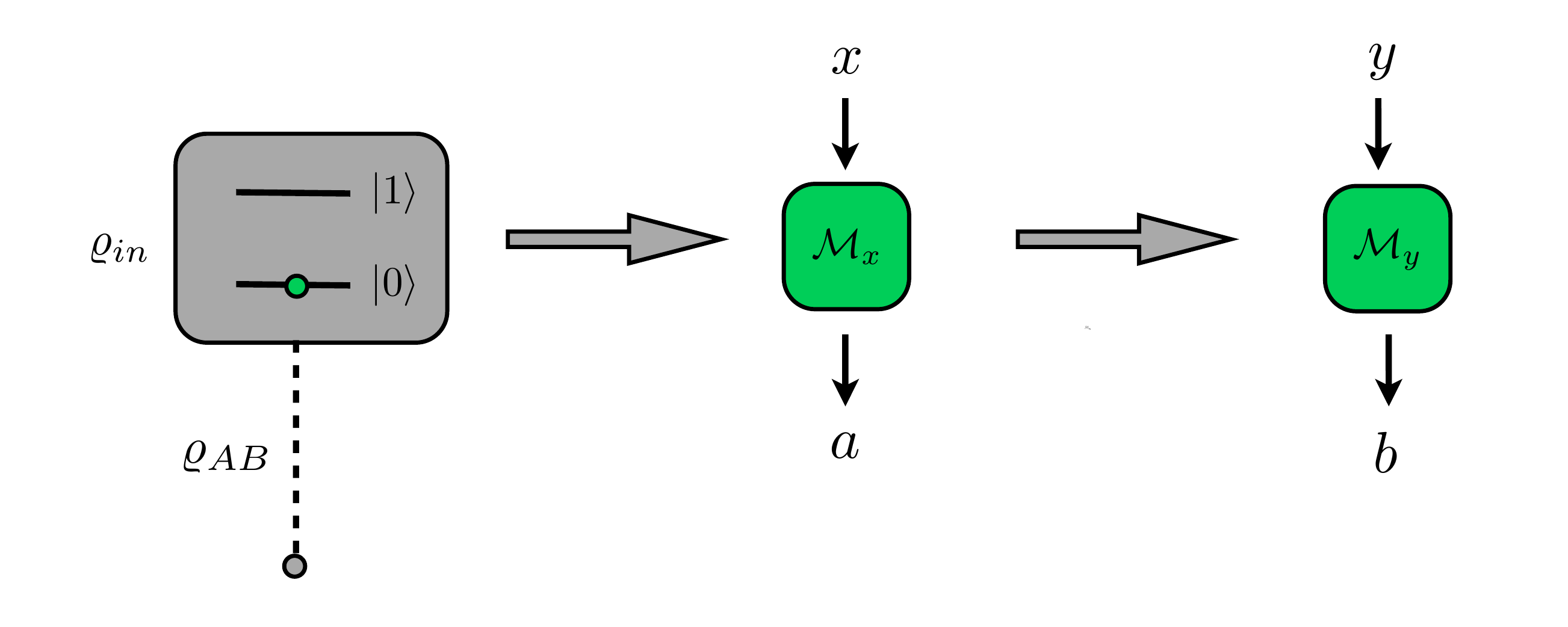} 
\end{center}
\caption{This figure shows schematically the scenario considered here. Sequences of measurements are performed on the qubit state $\varrho_{in}$, which may correspond to the reduced density matrix of $\varrho_{AB}$ which in turn describes a composite system. One observes temporal correlations $p(ab|xy)$ with measurement outcomes denoted by $a, b$ and measurement settings by $x, y$.}
\label{figschematic}
\end{figure}

{\it Using temporal correlations to obtain a lower bound on the purity.---}
We will  consider in the following sequences of general measurements acting on a single (sub)system whose (reduced) state is $\varrho_{in}$ (see Fig. \ref{figschematic}). To be more precise, we will examine the correlations $p(ab|xy)$ which correspond to the probability for obtaining outcome "$a$" in a first time step if one performs measurement "$x$" and then observing outcome "$b$" in a second time step if measurement "$y$" is performed. 

We will assume that one can use the same measurement apparatus at different time steps and the labeling of measurement settings does not change, e.g. in case $x=y$ one performs the same measurement twice, however, the outcomes do not need to be the same. The only further assumption will be that in the following the (sub)system that is measured is a two-dimensional system. In particular, we will not restrict the type of measurements, i.e. arbitrary instruments \footnote{Instruments are collections of completely positive maps (each associated to one outcome) for which the sum over all of these maps results in a trace-preserving map (see, e.g., \cite{instrument}).} are allowed. This scenario has been also considered in \cite{us},  \cite{othertheories} and \cite{expus}. 

We consider the following quantity:
\begin{align}
\BB_{1}&= p(\tiny{++}|00) +p(\tiny{++}|11) + p(\tiny{+-}|01) + p(\tiny{+-}|10).
\nonumber
\end{align}
It has been shown in \cite{us} that one can provide an (non-trivial) upper bound on $\mathcal{B}_1$ for general measurements on a qubit  which allows to employ $\mathcal{B}_1$ as a dimension witness. As we show here  it can also be used to witness the purity.

It can be proven that for any choice of measurements the maximum will be attained for pure initial and post-measurement states and that the maximum attainable value for fixed purity of the initial state will be monotonically increasing with increasing purity (see Appendix B). These relations are the key to use temporal correlations for obtaining lower bounds on the purity. In particular, it implies that in order to observe a certain value the system has to have at least a certain amount of purity. In Appendix A we will show that this key idea can in principle also be used to obtain lower bounds on the purity of the initial state for higher-dimensional systems and that it is essentially possible to employ $\mathcal{B}_1$ for this purpose. More precisely, we show  for arbitrary finite dimension that the maximal attainable value of any linear function of correlations of two time steps is a monotonically increasing function of the initial purity and that  the maximal value of $\mathcal{B}_1$ for arbitrary measurements  is not constant as a function of the purity.

For a qubit we provide here the explicit (analytic) relation between the maximal attainable value of $\BB_1$ and the purity. 
In order to ease the notation (and as it appears naturally in the derivations) we will from here on mainly refer to the length of the Bloch vector instead of the purity. That is we will use the Bloch decomposition for \bea \varrho_{in}=\frac{1}{2}(\one + p \,\vec{\alpha}_{in}\cdot \vec{\sigma})\eea with $\vec{\sigma}=(\sigma_x, \sigma_y,\sigma_z)$, $\sigma_i$ being the Pauli matrices, $0\leq p\leq 1$, $\vec{\alpha}_{in}\in \R^3$ and $|\vec{\alpha}_{in}|=1$. With this $p$ is the length of the Bloch vector and the purity of the initial state is given by   $\mathcal{P}(\varrho_{in})=1/2(1+p^2)$. Note that the purity $\mathcal{P}$ is monotonically increasing as a function of $p$ (and vice versa).
Let us then denote by $B_1(p)$   the maximal attainable value of $\BB_1$ for a given length of the Bloch vector, $p=\sqrt{2 \mathcal{P}-1}$,  of the initial state $\varrho_{in}$ and arbitrary choice of measurements. Then it holds  that \bea \label{Eqpinitial} B_1(p)=1/2(5+p).\eea
This relation follows from Th. \ref{Thupboundp} and we will discuss below how to derive it from this theorem.

The measurements that attain the maximum of $\BB_1$ are the same independent of the purity. In particular, the following protocol allows to attain $B_1(p)$. One of the measurement announces deterministically the outcome "+" and then prepares the state $1/2(\one - \vec{\alpha}_{in}\cdot \vec{\sigma})$. The other measurement measures the observable $\vec{\alpha}_{in}\cdot \nolinebreak\vec{\sigma}$. 

If one obtains in an experiment a value for $\BB_1$ denoted here and in the following by $\BB_1^{exp}$ one can straightforwardly deduce a lower bound on the purity of the measured initial state. This is due to the fact that  $B_1(p)$ is a monotonically increasing function of $\mathcal{P}$ [see Eq. (\ref{Eqpinitial})] and $B_1(p)\geq \BB_1^{exp}$ if the purity of the initial state that is measured in the experiment is given by $\mathcal{P}=1/2(1+p^2)$. The last relation captures that in an experiment the measurements that are implemented do not need to be the optimal ones that allow one to attain $\BB_1 (p)$. With this one obtains that in order to observe $\BB_1^{exp}$ a certain amount of purity is required. In particular, we obtain the following observation. 

\begin{observation}\label{obs1}
Let  $\BB_1^{exp}$ be the value for $\BB_1$ obtained in an experiment by performing sequences of measurements on the state $\varrho_{in}$. Then it holds for the purity $\mathcal{P}$ of $\varrho_{in}$ that
\begin{equation}\mathcal{P}\geq\frac{(2 \BB_1^{exp}-5)^2+1}{2}.\end{equation}
\end{observation}

Hence, temporal correlations allow one to witness the initial purity.

\begin{figure}[t!]
\begin{center}
\includegraphics[width=0.9\columnwidth]{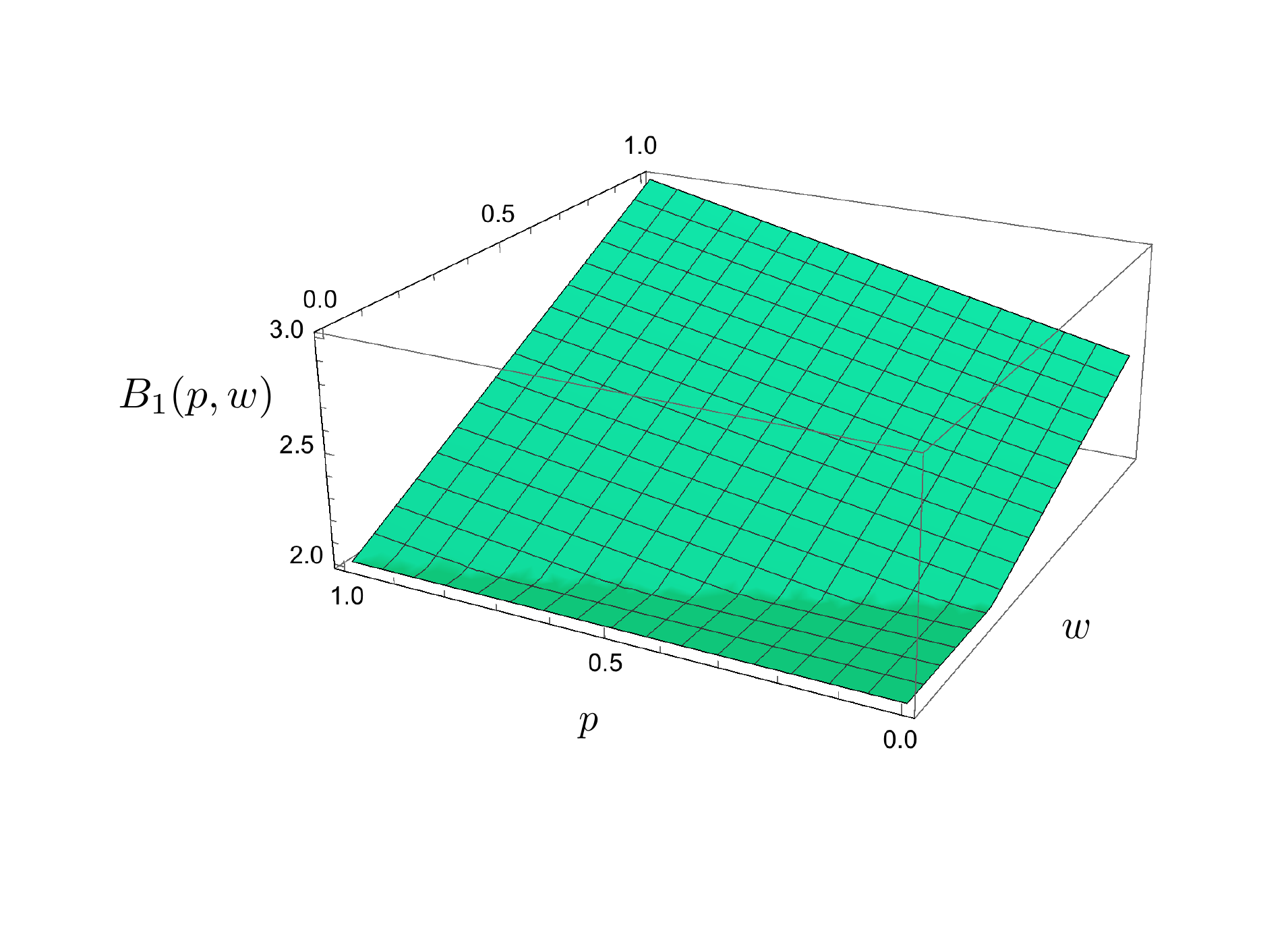} 
\end{center}
\caption{This figure shows the maximal attainable value of $\BB_1$ as function of a given Bloch vector length  of the initial state [i.e. purity $\mathcal{P}=1/2(1+p^2)$] and given Bloch vector length  of the post-measurements states for both measurements corresponding to outcome "+" [i.e. purity $\mathcal{W}=1/2(1+w^2)$], see Th. \ref{Thupboundp}.}
\label{fig:WP}
\end{figure}

Knowing the purity of the initial state it is also possible to deduce a lower bound on the maximal purity of the post-measurement state occurring at the second time step for outcome "+". To be more precise, one can provide a lower bound on the  state measured in the second time step, which here and in the following we will refer to  as post-measurement state. Let $p$ be the length of the Bloch vector of the initial state $\varrho_{in}$ and $w_{+|i}$ the one of the post-measurement state that is obtained after performing measurement $i\in\{0,1\}$ on $\varrho_{in}$ and observing outcome $"+"$.  Then one can determine the  maximum $B_1(p,w_{+|0}, w_{+|1})$ that is attainable with all measurements and states that respect the imposed purities. One can show that  $B_1(p,w_{+|0}, w_{+|1})$ is monotonically increasing as a function of $\mathcal{W}_{+|i}=1/2(1+w_{+|i}^2)$ (assuming the other purities fixed but arbitrary). Moreover, in an experiment leading to $\BB_1^{exp}$ in which the states occur with the respective purities it might be that one deviates from the optimal protocol. Hence, it holds for $w_{\max}=\max_{i\in\{0,1\}} w_{+|i} $ that 
\begin{align}\nonumber
B_1(p, w_{\max},w_{\max})\geq B_1(p, w_{+|0}, w_{+|1})\geq \BB_1^{exp}.
\end{align}
It only remains to determine $B_1(p, w)\equiv B_1(p, w, w)$ to provide an explicit lower bound on the maximal purity of the post-measurement states of outcome "+" depending on the purity of the input state. In the following theorem we provide a closed formula for $B_1(p, w)$ (see also Fig. \ref{fig:WP}).
\begin{theorem}\label{Thupboundp}
Let $\mathcal{P}$ be the purity of the initial state and $\mathcal{W}$ the purity of the post-measurement states that occur for measurement $i\in\{0,1\}$ observing outcome "+". Then for a two-dimensional system the maximal value of $\BB_1$, $B_1(p, w)$, that can be obtained for arbitrary initial states and measurements that respect these constraints on the purities, is given by 
\bea \nonumber B_1(p, w)=\begin{cases} 
     2 &  0\leq w\leq  \frac{1-p}{3+p}\\
1 + \frac{1 + w}{2} + \frac{(1 + p)(1 + w)}{4} &  \frac{1-p}{3+p}< w\leq 1 \\
     \end{cases},\eea
where $w=\sqrt{2 \mathcal{W}-1}$ and $p=\sqrt{2 \mathcal{P}-1}$ are the length of the Bloch vector for the respective purity.
\end{theorem}
The proof of this theorem can be found in Appendix B. 
This theorem allows one to deduce a lower bound on the maximal purity of the post-measurement states provided that the purity of the initial state is known. In particular, we have that for $ \BB_1^{exp}\leq 2$ we cannot deduce a lower bound, however if $ \BB_1^{exp}> 2$ it follows from the Theorem above and $B_1(p, w_{\max})\geq \BB_1^{exp}$  that 
\begin{align}\nonumber\mathcal{W}_{\max}&\geq \frac{14 + 4 (\BB_1^{exp})^2 + \mathcal{P} + 5 \sqrt{2 \,\mathcal{P}-1}}{4 + \mathcal{P} + 3 \sqrt{2 \,\mathcal{P}-1}}\\ &-\frac{
 2 \BB_1^{exp} (7 + \sqrt{ 2\, \mathcal{P}-1})}{4 + \mathcal{P} + 3 \sqrt{2 \,\mathcal{P}-1}}.\end{align}
Moreover, note that as $B_1(p, w)$ is monotonically increasing as a function of $\mathcal{W}$ we also have that  \begin{align}\nonumber B_1(p)&=\max_{0\leq w\leq 1}B_1(p, w)\\&=B_1(p,1)=B_1(p)=\frac{5+p}{2},\end{align}
which allows to bound the purity of the initial state as argued above [see Eq. (\ref{Eqpinitial}) and Observation \ref{obs1}]. 

In an experiment one may not be able to perfectly implement the measurements that realize the tight bound in Eq. (\ref{Eqpinitial}) but one may have post-measurement states which are not perfectly pure. Theorem  \ref{Thupboundp} allows to deduce how robust the estimation of the initial purity is with respect to not perfectly pure post-measurement states as it specifies also for this case  the maximal attainable value. Consider the case that the maximal length of the Bloch vectors of the post-measurement state is given by $1-\epsilon$. Then it holds that $ B_1(p,1-\epsilon)=B_1(p)-\frac{3+p}{4}\epsilon $, i.e the deviation is at most $\epsilon$ and also linear in this parameter. Note further  that in case one can estimate a lower bound on the purity of the post-measurement state one can use Theorem \ref{Thupboundp} also to improve the lower bound on the purity of the initial state (by using the lower bound for the respecitive $w$ in Theorem \ref{Thupboundp}). As this requires information about the measurement the so obtained bound is not semi-device independent anymore. 

{\it Upper bound on the concurrence based on the purity.---}
As mentioned before it is well known that for bipartite pure states there is a close connection between entanglement and the purity of the reduced state of a single party. In particular, the reduced state is pure only for product states, whereas for maximally entangled states it is maximally mixed. For mixed states and on a more quantitative level, entanglement measures such as the concurrence are defined as the convex roof extension of a function of the local purity. More precisely, the concurrence \cite{conc, conc2}  is given by
\begin{align}\label{convexroof}
C(\rho)=\inf \sum_i q_i C( \ket{\psi_i}),
\end{align}
where the infimum is taken over all pure state decompositions, $\rho= \sum_{i} q_i \proj{\psi_i}$, and $C( \ket{\psi_i})=\sqrt{ 2\{1-\tr[ (\rho_A^i)^2]\}}$ with $\rho_A^i=\tr_B(\proj{\psi_i})$. It seems therefore natural to consider the relation among the concurrence and the purity of the reduced state more closely in order to obtain a bound on the concurrence. The following result  will allow us to provide a upper bound on the concurrence based on the observed temporal correlations. For two-qubit states $\varrho_{AB}$ with $\rho_A=\tr_B(\varrho_{AB})$ and  $\rho_B=\tr_A(\varrho_{AB})$ it holds that \cite{Concbound} \bea C(\varrho)\leq \min_{X\in\{A, B\}}\sqrt{2\{1-\tr [(\varrho_X)^2]\}}.\eea
This bound has been already observed for arbitrary bipartite $d$-dimensional states in \cite{Concbound}. For completeness we will nevertheless present in Appendix C  a (alternative but similar) proof for two-qubit states.
Combining this with the lower bound on the purity based on temporal correlations (see Observation \ref{obs1}) we can state the following observation. 

\begin{observation} 
Let $\varrho_{AB}$  be a two-qubit state and $\BB_1^{exp}$ the experimental value for $\BB_1$ obtained for sequences of measurements on one of the subsystems. Then it holds for the concurrence $C(\varrho_{AB})$ that
\begin{equation}C(\varrho_{AB})\leq\sqrt{1-(2 \BB_1^{exp}-5)^2}.\end{equation}
\end{observation}

Moreover, it has been also shown in \cite{Concbound} that for multipartite states $C(\varrho)\leq 2^{1-n/2}\sqrt{2^{n}-2-\sum_i \tr[ (\varrho_i)^2]\}}$. Here $C(\varrho)$ is a generalization of the concurrence to the multipartite case  defined by $C(\psi)=2^{1-n/2} \sqrt{2^{n}-2-\sum_i \tr[ (\varrho_i)^2]\}}$ \cite{concmulti1,concmulti2}, where $n$ is the number of parties, $\varrho_i$ are the single-party density matrices, and $C(\varrho)$ is obtained via the convex roof extension from $C(\psi)$ (see Eq. (\ref{convexroof})).

Hence, also for multipartite system one can first obtain from the correlations that arise from sequences of local measurements on subsystems a semi-device-independent lower bound on the purity of the subsystems and with this then an upper bound on the concurrence of the joint system.

{\it Summary and outlook.---}
In this work we considered sequential measurements on a qubit. We showed that one can deduce from the observed correlations a lower bound on the purity of the initial state of the qubit. In case the qubit is part of a two-qubit system, this provides an upper bound on the concurrence. Moreover, provided that the purity of the initial state is known our approach allows one to obtain a lower bound on the maximal purity of the post-measurement states occuring at the second time step for one of the outcomes. Our result shows that it is possible to use temporal correlations for bounds on the purity and the concurrence by considering explicitely the example of a qubit. Moreover, we proved that also for higher-dimensional systems it is essentially possible to employ temporal correlations in order to establish bounds on the initial purity. It would be relevant to pursue our investigation of higher-dimensional systems and provide explicit purity witnesses. Moreover, it would be interesting to see whether longer sequences allow  in principle for a better performance as has been observed for the case of dimension witnesses \cite{expus}.

I thank Otfried G\"uhne, Costantino Budroni and Nikolai Wyderka for useful discussions. Moreover, I would like to thank Otfried G\"uhne  for careful reading of the manuscript. This work has been supported by the the Austrian Science Fund (FWF): J 4258-N27, the ERC (Consolidator Grant 683107/TempoQ) and the DAAD (Projekt-ID: 57445566).


\appendix
\section{Temporal correlations allow to witness the purity for $d$-dimensional systems\label{Apphighd}}
In this Appendix we show that it is essentially possible to use temporal correlations for providing lower bounds on the purity for $d$-dimensional systems. In particular, we will prove that one can construct functions of the correlations whose maximum for arbitrary measurements is monotonically increasing as a function of the purity of the initial state.  Hence, in order to observe a certain value of this function one has to have a certain amount of purity and therefore these can be used to provide lower bounds on the purity. Moreover, we will show that in principle the quantity $\mathcal{B}_1$ could also be used to gain information about the the purity for $d$-dimensional systems.
\begin{proof}
In order to do so, we consider a quantity $\mathcal{R}=\sum \alpha_{a b x y} p(a b|x y)$, which is linear in the correlations and therefore also in the initial state. More precisely, it holds that $ p(a b|x y)= p(a|x) p(b|axy)= \tr (\mathcal{E}_{a|x}\varrho_{in}) p(b|axy)$ with $\mathcal{E}_{a|x}$ being the effect for the measurement in the first time step. As we will here and in the following restrict to two time steps and furthermore be interested in the optimal measurements it is possible to use as description of the measurements their effects and post-measurement states (instead of describing them via instruments).  The effects  $\mathcal{E}_{a|x}$ are positive semi-definite matrices with the property that $\sum_a \mathcal{E}_{a|x}=\one$. They allow to calculate the probability for obtaining outcome "a" by implementing measurement x on a state $\rho$ via $\tr (\rho \mathcal{E}_{a|x})$. Note that for any combination of effects and post-measurement states there exists a valid instrument that realizes them when being applied to an arbitrary $\varrho_{in}$. This can be achieved by first applying an instrument with the desired effects and then preparing the system in the desired post-measurement state. Note that in case more time steps are considered not any sequence of post-measurement states is possible. However, note that by considering the description in terms of instruments and  the Heisenberg picture  the following argumentation can be straightforwardly generalized to longer sequences.

Now let us assume that for a given purity of the initial state $\mathcal{P}$ one knows the optimal protocol that maximize $\mathcal{R}$. We then have $\mathcal{P} =\sum_i q_i^2$ with $q_i$ being the eigenvalues of the optimal $\varrho_{in}$.  Let us denote the eigenbasis for $\varrho_{in}$ by $\{\ket{i}\}$, the corresponding optimal effects by $\tilde{\mathcal{E}}_{a|x}$ and the maximal attainable value by $R (\mathcal{P})$.  Then one obtains that 
\begin{equation}\label{AppA1}R (\mathcal{P})= \sum_{i, a, x, b, y} q_i \alpha_{a b x y} \tr (\tilde{\mathcal{E}}_{a|x} \proj{i}) p(b|axy).\end{equation} Note that there always exists a state $\ket{k}$ in the eigenbasis for which 
\begin{align}\label{AppA2}
\sum_{a, x, b, y} \alpha_{a b x y} \tr (\tilde{\mathcal{E}}_{a|x} \proj{k}) p(b|axy)\geq\\\nonumber \sum_{a, x, b, y} \alpha_{a b x y} \tr (\tilde{\mathcal{E}}_{a|x} \proj{j}) p(b|axy)
\end{align}
 for all $\ket{j}$. Note further that as we assume the optimal strategy one can choose wlog that $q_k\geq q_j$. This is due to the fact that if $q_k< q_j$ for some $\ket{j}$ for which the inequality in Eq. (\ref{AppA2}) is strict one could apply a unitary exchanging $\ket{k}$ and $\ket{j}$ before and after the supposedly optimal measurements, and obtain a higher value for $\mathcal{R}$ which contradicts our assumption that we are implementing the optimal protocol. In case the inequality is an equality we can simply relabel $\ket{k}$ to obtain $q_k\geq q_j$. 
 
It then remains to show that  by increasing the purity it is possible to increase the maximal value of $\mathcal{R}$. Let us first consider the case that the inequality in Eq. (\ref{AppA2}) is strict for at least one $\ket{j}$ with $q_j\neq 0$ which we will denote by $\ket{l}$.  Then for any purity $\mathcal{Q}>\mathcal{P}$ one can find a value $\epsilon>0$ such that $\mathcal{Q}= (q_k+\epsilon)^2+(q_l-\epsilon)^2+\sum_{i\neq k, l} q_i^2$, i.e.  $\epsilon=\frac{q_l-q_k+\sqrt{2 (\mathcal{Q}-\mathcal{P})+(q_k-q_l)^2}}{2}$. We will then use the notation $\tilde{q}_i=q_i$ for $i \not\in \{k, l\}$, $\tilde{q}_k=q_k+\epsilon$ and $\tilde{q}_l=q_l-\epsilon$. It is then straightforward to see that choosing the same effects and imposing the same post-measurement states as before (e.g. by considering some measure-and-prepare channel) one obtains that
\begin{equation}
\sum_{a, x, b, y, i} \tilde{q}_i \alpha_{a b x y} \tr (\tilde{\mathcal{E}}_{a|x} \proj{i}) p(b|axy)>R(\mathcal{P}). \end{equation}
 Hence, we have shown that for any purity $\mathcal{Q}>\mathcal{P}$ there exists a strategy (measurements) that allow to exceed the maximal attainable value $R(\mathcal{P})$ in case inequality  (\ref{AppA2}) is strict for at least one $\ket{j}$ for which $q_j\neq 0$.

 Note that in case the inequality   (\ref{AppA2}) is an equality for some set $j\in\mathcal{J}=\{j_1,\ldots, j_n\}$ and $q_i=0$ for $i\not\in \mathcal{J}$ it can be easily seen that this implies that for any purity which can be realized with a density matrix of rank $n$ or smaller the value $R(\mathcal{P})$ can be attained. Note further that this implies that for any purity $\mathcal{Q}$ which corresponds to a density matrix $\rho_n$ of rank $n$ and $\mathcal{Q}<\mathcal{P}$ the optimal strategy has the property that for the whole eigenbasis (with non-zero eigenvalue) of $\rho_n$ one obtains an equality as otherwise the maximal attainable value of $\mathcal{R}$ has to strictly increase with increasing purity as we just have shown. However, this implies that also with higher purity this value is attainable and therefore it has to hold that in this case we have that $R(\mathcal{Q})=R(\mathcal{P})$.  Increasing now $\mathcal{P}$ one obtains that  the maximal attainable value is at least $R(\mathcal{P})$ and either remains constant or increases by the argumentation before. 
 
In summary we have that for both cases the maximal attainable value of $\mathcal{R}$ is monotonically increasing as a function of the purity. Hence, in case it is not constant for all purities $\mathcal{R}$ can be used to obtain some information on the purity. It is obvious that for any dimension $d$ there exists some quantity $\mathcal{R}$ whose maximum for given purity does not remain constant for all purities. 
 
As an example consider $\mathcal{B}_1$ for which one can show that for the maximally mixed state the maximum is upper bounded by  $\max [3, 4(1-1/d)]$ but the maximal value for a pure state corresponds for $d\geq 3$ to $4$. This implies that also for higher dimensions the maximal attainable value of $\mathcal{B}_1$ is not constant as a function of the purity. 
In order to see the upper bound on $\mathcal{B}_1$ for the maximally mixed state note first that one can use an analogous argumentation to before to show that for fixed purity of the initial state either the maximal attainable value of $\mathcal{B}_1$ is monotonically increasing as a function of the purity of the post-measurement states. Hence,  the optimal post-measurement states are either pure or can be chosen to be pure. As then there are only two pure post-measurement states appearing in the quantity this implies that only a two-dimensional subspace is relevant for the measurements in the second time step. Moreover, considering the first time step it is then straightforward to see  that in order to obtain the maximum the diagonal terms in the effects for outcome "+" should be one in the orthogonal complement to this subspace and terms mixing the qubit subspace and its complement are chosen to be zero (in order to not introduce further constraints on the two-dimensional subspace due to positivity). We then use that one can parametrize the restriction of the effects to the two-dimensional subspace and the states as in \cite{us}. That is one can use for such effects the parametrization $\mathcal{E}_{+|i}=a_i(\one_2 +b_i \vec{\sigma}\cdot\vec{c}_i)+ \one_{d-2}$ where $\vec{c}_i\in \R^3$, $|\vec{c}_i|=1$, $0\leq a_i \leq 1/(1+b_i)$, $0\leq b_i\leq 1$,   $\one_x$ denotes the $x$-dimensional identity and  $\vec{\sigma}$  ($\one_2$)  the vector of 
Pauli matrices (the identitiy) in the qubit subspace, respectively. Using then that the initial state is maximally mixed one can show analogously to \cite{us} that the maximum of $\mathcal{B}_1$ is smaller or equal to $\max [3, 4(1-1/d)]$ or it is attained when the effects are projective.  More precisely, consider first the points where the derivative with respect to $a_0$ (assuming all other parameter to be fixed but arbitrary) vanishes, i.e.
\begin{align}\nonumber
a_0\frac{d \BB_1}{d a_0}&=\left[p(+|0)-\frac{d-2}{d}\right][p(\tiny{+}|+00) + p(\tiny{-}|+01)]\\\nonumber
&+ p(+|1)[p(\tiny{-}|+10)-1]+p(+|0)p(\tiny{+}|+00)=0,
\end{align}
where we used $p(+b|xy)=p(+|x) p(b|+xy)$.
Hence, at the points where the derivative vanishes it holds that 
\begin{align}\nonumber
&p(+|0)[p(\tiny{+}|+00) + p(\tiny{-}|+01)]+ p(+|1)[p(\tiny{-}|+10)=\\\nonumber&p(+|1)+\frac{d-2}{d}[p(\tiny{+}|+00) + p(\tiny{-}|+01)]-p(+|1)p(\tiny{+}|+00)\\&\leq 3-\frac{4}{d}\label{derivatprop1}
\end{align}
and therefore $\BB_1\leq 4(1-1/d)$ at these points. It remains to consider the boundary $a_0=0$ and $a_0 = 1/(1+b_0)$. It is easy to see that for $a_0=0$ it holds that $\BB_1\leq 2$. Before considering the case $a_0 = 1/(1+b_0)$ note that the quantity $\BB_1$ is symmetric with respect to the exchange of measurement setting 0 and 1. Hence,  in order to possibly achieve a value for $\BB_1$ that is greater than $4(1-1/d)$ measurements with  $a_1 = 1/(1+b_1)$ have to be used. As can be easily seen the choice $a_i = 1/(1+b_i)$ corresponds to measurements for which the effect corresponding to the outcome "-" is proportional to a projector, i.e. 
\begin{equation}\label{propproj}
\mathcal{E}_{-|i}=\frac{u_i}{2}(\one_2 - \vec{\sigma}\cdot\vec{c}_i).
\end{equation}
with $0\leq u_i\leq 1$.
One can then use exactly the same argumentation as in Appendix C.1. of \cite{us} to show that either $\BB_1\leq 3$ or the effects have to be projective, i.e. one shows that at the points where the derivative with respect to $u_i$ vanishes it holds that  $\BB_1\leq 3$ and that the same holds true for the boundary point $u_i=0$.
Considering then projective effects and the optimal choice of post-measurement states for such effects as in Appendix C.1. of \cite{us} the quantity depends on one remaining parameter, the angle between the Bloch vectors in the restriction to the two-dimensional subspace of the two measurements. It is then straight forward to see that the maximum attainable value with projective effects is given by $4(1-1/d)$ which is obtained for $\vec{c}_0=-\vec{c}_1$. In summary, we have shown that for the maximally mixed state it is not possible to exceed $\max [3, 4(1-1/d)]$. In particular, for $d\geq 4$ we have that $4(1-1/d)\geq 3$ and in this case the bound can be reached. For pure initial states one can attain a value of $4$ in case $d\geq 3$ (see \cite{us}). Hence, we have that the maximal attainable value  $\mathcal{B}_1$ is not constant but due to the argumentation above at least on some interval(s) strictly increasing with increasing purity. This concludes the proof that temporal correlations can be used to build witnesses for the purity of $d$- dimensional states. 
 \end{proof}
\section{Proof of Theorem  2\label{AppTC}}
In this appendix we will first show that $B_1(p, w_{+|0},w_{+|1})$ (as defined in the main text and below) is monotonically increasing as a function of $w_{+|1}$ (and therefore also $\mathcal{W}_{+|i}$). Moreover, we will prove Theorem 2.

Recall first that $B_1(p, w_{+|0},w_{+|1})$ is the maximal value for $\BB_1$ that is attainable with arbitrary (time-independent) measurements for a given purity $\mathcal{P}=1/2(1+p^2)$ of the initial state and fixed purity of the states that are measured at the second time step $\mathcal{W}_{+|i}=1/2(1+w_{+|i}^2)$ if in the first time step measurement $i$ is performed and outcome $"+"$ is obtained. Analogously to the proof in Appendix A one can show in general that for two time steps the maximum (with respect to all other parameters but the purities of the states which are assumed to be fixed apart from the purity of one post-measurement state) of any linear function of  temporal correlations  is also a monotonically increasing function of the purity for the post-measurement states. This implies that in particular $B_1(p, w_{+|0},w_{+|1})$ is monotonically increasing as a function of $\mathcal{W}_{+|i}$  (assuming that all other parameters are fixed but arbitrary).

Alternatively one can also show that  $B_1(p, w_{+|0},w_{+|1})$ is monotonically increasing as a function of $\mathcal{W}_{+|i}$ by considering the derivative with respect to $w_{+|i}$. 

In order to do so we parametrize the effects via $\mathcal{E}_{+|x}=r_x \one + q_x \vec{v}_x\cdot \vec{\sigma}$ and $ \mathcal{E}_{-|x}=\one -\mathcal{E}_{+|x}$ for $x\in\{0,1\}$ with $0\leq q_x\leq r_x\leq 1-q_x$, $\vec{v}_x\in \R^3$,  $|\vec{v}_x|=1$ and $\vec{\sigma}$ is a vector containing the Pauli matrices. As mentioned in the main part of the manuscript one can use the Bloch decomposition to parametrize states with fixed purity, i.e. \bea\label{bloch}\varrho=1/2(\one + w \vec{\alpha}\cdot \vec{\sigma}),\eea with $|\vec{\alpha}|=1$ and $0\leq w\leq 1$. The purity $\mathcal{W}$ is  then related to the length of the Bloch vector $w$ via 
\bea\mathcal{W}=\frac{1}{2}(1+w^2).\eea
Using this parametrization for the states one can analogously as in Appendix C of \cite{us} determine the initial and post-measurement states that maximize $\BB_1$ for given effects and purities. More precisely, the  Bloch vectors for the post-measurement states of measurement $i\in\{0,1\}$ are proportional to $(-1)^i (q_0  \vec{v}_0-q_1 \vec{v}_1)$ and the Bloch vector for the initial state has to be chosen proportional to $q_0 X_0  \vec{v}_0+q_1 X_1  \vec{v}_1$  with $X_0=1+r_0-r_1+w_{+|0}\sqrt{q_0^2+q_1^2-2 q_0q_1\vec{v}_0\cdot\vec{v}_1}$  and $X_1=1+r_1-r_0+w_{+|1}\sqrt{q_0^2+q_1^2-2 q_0q_1\vec{v}_0\cdot\vec{v}_1} $. This is due to the fact that the choice of $\vec{\alpha}$ which maximizes  $\vec{\alpha}\cdot \vec{\beta}$ under the constraint that the length of $\vec{\alpha}$ is fixed is given by $ \vec{\alpha}=c \vec{\beta}$ with $c\geq 0$ and $c$ ensuring the correct length of the vector. For the optimal choice of states (and arbitrary effects) one observes that $\BB_1$ is a linear function of $p$ and one can show that $d\BB_1/dw_{+|i}\geq 0$. Hence, $\BB_1$ is monotonically increasing as function of $w_{+|i}$ which implies that it is also a monotonically increasing function of $\mathcal{W}_{+|i}$. In particular, we have that \begin{align}\nonumber
B_1(p, w_{\max}, w_{\max})\geq B_1(p, w_{+|0}, w_{+|1}),
\end{align}
where $ w_{\max}=\max_{i\in\{0,1\}} w_{+|i}$.\\

In the following we will use the notation $B_1(p, w)\equiv B_1(p, w, w)$.
We will next show Theorem 2. In order to improve readability we repeat the theorem here.\\ \\
\noindent{\it{\bf Theorem 2. }Let $\mathcal{P}$ be the purity of the initial state and $\mathcal{W}$ the purity of the post-measurement states that occur for measurement $i\in\{0,1\}$ observing outcome "+". Then for a two-dimensional system the maximal value of $\BB_1$, $B_1(p, w)$, that can be obtained for arbitrary initial states and measurements that respect these constraints on the purities, is given by 
\bea \nonumber B_1(p, w)=\begin{cases} 
     2 &  0\leq w\leq  \frac{1-p}{3+p}\\
1 + \frac{1 + w}{2} + \frac{(1 + p)(1 + w)}{4} &  \frac{1-p}{3+p}< w\leq 1 \\
     \end{cases},\eea
where $w=\sqrt{2 \mathcal{W}-1}$ and $p=\sqrt{2 \mathcal{P}-1}$ are the length of the Bloch vector for the respective purity.}
\begin{proof}
Note first that by deterministically assigning outcome "+" for both measurements  independent of the state that is measured one obtains that $\BB_1=2$. Moreover, by using the following protocol one can obtain  $\BB_1=1 + \frac{1 + w}{2} + \frac{(1 + p)(1 + w)}{4}$. Let the initial state have a Bloch vector of length $p$ pointing in z-direction, i.e. in Eq. (\ref{bloch}) we have that $\vec{\alpha}=(0,0,1)$. One of the measurement is chosen to be of the form that one deterministically announces "+" and prepares the state with Bloch vector pointing in - z-direction, i.e. $\vec{\alpha}=(0,0,-1)$, and of length $w$. The other measurement performs a projective measurement in the computational basis $1/2(\one\pm \sigma_z)$ with associated outcome "$\pm$" and then prepares the state with $\vec{\alpha}=(0,0,1)$ and length $w$.
Hence, the values for $\BB_1$ given in the theorem above are attainable. Moreover, note that 
$2\geq 1 + \frac{1 + w}{2} + \frac{(1 + p)(1 + w)}{4}$ if and only if $w\leq  \frac{1-p}{3+p}$. It remains to show that $\BB_1$ for given $p$ and $w$ cannot exceed $\max [2, 1 + \frac{1 + w}{2} + \frac{(1 + p)(1 + w)}{4}]$.

In order to do so, we note first that it can be easily seen using the same argumentation \footnote{In \cite{us} we argued that either the maximum of $\BB_1$ is smaller or equal than 3 or the effects for each measurement corresponding to outcome "-" have rank 1. Examining the points where the derivative with respect to $a_i$ vanishes more closely, it can be easily seen that $\BB_1\leq 2$ at that points for any given states (as is explained in the following).}  as in \cite{us} that either $B_1(p, w)\leq 2$ or for both measurements the effects for outcome "-" are proportional to projectors. That is one uses the parametrization  
\begin{equation}\label{eff0}\mathcal{E}_{+|i}=a_i(\one_2 +b_i \vec{\sigma}\cdot\vec{c}_i)
\end{equation}
with $0\leq a_i \leq 1/(1+b_i)$, $0\leq a_i\leq 1$ and $\vec{c}_i$  being a real vector of unit length and considers the critical points with respect to $a_i$ (all other parameter are assumed to be fixed but arbitrary). One observes that at the points where the derivative vanishes Eq. (\ref{derivatprop1}) with $d=2$ holds. Hence, at these points $\BB_1\leq 2$. It remains to consider the boundary of the interval $0\leq a_i \leq 1/(1+b_i)$. For $a_i=0$ one obtains also $\BB_1\leq 2$. For $a_i = 1/(1+b_i)$ the effect corresponding to outcome "-" is proportional to a projector. 

Using then the parametrization of effects as given in Eq. (\ref{propproj}) and considering the points where the gradient with respect to $u_0$ and $u_1$ (assuming again all other parameters to be fixed but arbitrary) vanishes, we obtain that \bea \sum_i u_i \frac{\partial \BB_1}{\partial u_i}=0.
\eea
This is equivalent to
\begin{align}\nonumber\BB_1&=\frac{1}{2}[p(+|0)+p(+|1)+p(+|+00)\\\label{pq}
&+p(-|+01)+p(+|+11)+p(-|+10)],
\end{align}
where we used that one can write $p(+b|xy)=p(+|x) p(b|+xy)$. By maximizing the right hand side of this equation one can obtain an upper bound on $\BB_1$ at the points where the gradient vanishes. Note first that the expression is a linear function in the parameters $u_i$ and therefore is maximal at one of the boundary points $u_i=0$ or $u_i=1$. If $u_0=u_1=0$ then independent of the measured states outcome "-" never occurs and therefore the right hand side is upper bounded by two. In case $u_0=u_1=1$ the effects are projective and choosing the optimal initial and post-measurement states analogous to \cite{us} we get for the right hand side
\bea\label{rhs1}
\frac{1}{4}[6+ 2 w \sqrt{2-2 x} +p \sqrt{2+2 x}],
\eea
where here and in the following $x$ corresponds to the angle between the Bloch vectors of the effects for outcome "+" of the two measurements, i.e. $\vec{c}_0\cdot \vec{c}_1=x$. More precisely, one chooses the vector $\vec{\alpha}$ in Eq. (\ref{bloch}) for the initial state proportional to $\vec{c}_0+\vec{c}_1$ and for the post-measurement states proportional to $\pm(\vec{c}_0-\vec{c}_1)$. This choice is optimal, as the maximal value of an expression of the form $\vec{\alpha}\cdot \vec{v}$ is attained if $\vec{\alpha}$ is chosen parallel to $\vec{v}$. One can then easily show that Eq.(\ref{rhs1}) is maximized for the point where the derivative with respect to $x$ vanishes (and not at the boundary given by $x\in\{\pm 1\}$), i.e. $x = (p^2 - 4 w^2)/(p^2 + 4 w^2)$. This results in a maximal value for the right hand side that is strictly smaller than $1 + \frac{1 + w}{2} + \frac{(1 + p)(1 + w)}{4}$ for all possible values of $p$ and $w$. It remains then to consider $u_0=0$ and $u_1=1$ as Eq. (\ref{pq}) is symmetric with respect to the exchange of measurement 0 and 1. It can be easily seen that in this case the right hand side of Eq. (\ref{pq}) is at most
\bea
\frac{1}{2}[3+ \frac{1 + p}{2} +w]\leq 1 + \frac{1 + w}{2} + \frac{(1 + p)(1 + w)}{4}.
\eea
In summary we have seen that for the points where the gradient with respect to $u_i$ vanishes $\BB_1\leq 2$ or $\BB_1\leq 1 + \frac{1 + w}{2} + \frac{(1 + p)(1 + w)}{4}$ for the given purities which implies in particular that $\BB_1\leq \max (2, 1 + \frac{1 + w}{2} + \frac{(1 + p)(1 + w)}{4})$. In order to prove the theorem it therefore remains to show that this upper bound also holds true at the boundary of the domain $0\leq u_i\leq 1$, i.e. the effect for one of the measurements is either projective [case A] or the identity [case B]. Note that $\BB_1$ is symmetric regarding the exchange of the measurements. Let us first discuss case A and choose without loss of generality $u_1=1$, i.e. measurement 1 is projective. At the points where the derivative with respect to $u_0$ vanishes one obtains that 
\begin{align}\nonumber\BB_1&=\{[p(+|0)-1][1-p(+|+00)]+1+\\\nonumber
&+p(-|+01)+p(+|1)p(+|+11)\}\\&\leq 1 + \frac{1 + w}{2} + \frac{(1 + p)(1 + w)}{4}
\end{align}
 for all $p$ and $w$ if $u_1=1$. It remains to consider for case A the boundary points $u_0=0$ and $u_0=1$. The case $u_0=0$ corresponds to a deterministic assignment of outcome "+" and is included in case B. Choosing analogous to before (see also \cite{us}) the optimal states for the measurements with $u_0=u_1=1$  one obtains that in this case
 \bea
\frac{1}{4}(2+ w \sqrt{2-2 x})(2 +p \sqrt{2+2 x}).
\eea
It can be shown that at the critical points this function is smaller or equal to $1 + \frac{1 + w}{2} + \frac{(1 + p)(1 + w)}{4}$ and therefore with projective effects one cannot exceed this value. We will proceed with case B and choose without loss of generality $u_0=0$. It is then immediate to see that the Bloch vectors of the optimal states have to be chosen parallel or antiparallel to the Bloch vector of measurement 1. For this choice of states and measurements one obtains that 
\bea\nonumber
\BB_1=\frac{1}{4} [8 + (1 - p)  (1 - w)u_1^2 + 2 u_1 (-1 + p + 2 w)].
\eea
It can be checked that for the boundary points $u_1=0$ and $u_1=1$ this implies that $\BB_1=2$ and $\BB_1=1 + \frac{1 + w}{2} + \frac{(1 + p)(1 + w)}{4}$. Moreover, it can be easily seen that the point where the derivative with respect to $u_1$ vanishes corresponds to a minimum. In summary, this concludes the proof that $\BB_1$ for given length of the Bloch vectors of the states, $w$ and $p$, is upper bounded by $\max (2, 1 + \frac{1 + w}{2} + \frac{(1 + p)(1 + w)}{4})$. Recall that this bound is tight and that $2\geq 1 + \frac{1 + w}{2} + \frac{(1 + p)(1 + w)}{4}$ if and only if $w\leq  \frac{1-p}{3+p}$, which proves the theorem.
 \end{proof}

\section{Proof of the upper bound on the concurrence based on the purity of a subsystem\label{AppConc}}
It should be noted that the upper bound on the concurrence given by $C(\varrho_{AB})\leq \min_{X\in\{A, B\}}\sqrt{2\{1-\tr [(\varrho_X)^2]\}}$ has already been proven for arbitrary bipartite $d$-dimension systems in \cite{Concbound}.
For the sake of completeness we provide here a (alternative but similar) proof for two-qubit states.
\begin{proof}
We will use in the following that in the two-qubit case it has been proven in \cite{Wootters} that for any $\varrho$ there exists some decomposition into pure states, $\varrho=\sum_{i=1}^r p_i \proj{\phi_i}$, such that $C(\varrho)=C( \ket{\phi_i})$ for all $i\in\{1,\ldots,r\}$. Moreover, recall that  it holds for the pure states $\ket{\phi_i}$ that $C( \ket{\phi_i})=\sqrt{2\{1-\tr [(\varrho_A^i)^2]\}}$ with $\varrho_A^i=\tr_B(\proj{\phi_i}) $. Note that due to  $C( \ket{\phi_i})=C( \ket{\phi_j})$ we have therefore \bea \label{eq} \tr [(\varrho_A^i)^2]=\tr [(\varrho_A^j)^2]\equiv Q(\varrho).\eea From this equation it follows that \begin{align}
\tr [(\varrho_A)^2]&=\sum_{i,j} p_i p_j \tr(\varrho_A^i\varrho_A^j) \\\nonumber
&\leq \sum_{i,j} p_i p_j \sqrt{\tr [(\varrho_A^i)^2]}\sqrt{\tr [(\varrho_A^j)^2]}\\\nonumber
&=\sum_{i,j} p_i p_j \tr [(\varrho_A^i)^2]=Q(\varrho).
\end{align}
The inequality arises from the Cauchy-Schwarz inequality (using the Hilbert-Schmidt inner product  for each summand) and then we use Eq. (\ref{eq}) and $\sum p_i=1$. Hence, we have that \begin{align}C(\varrho)=C( \ket{\phi_i})=\sqrt{2[1-Q(\varrho)]}\leq \sqrt{2\{1-\tr [(\varrho_A)^2]\}}.\end{align} One can show analogously that the bound also holds true for $\varrho_B$ which proves the statement.\end{proof}



\begin{thebibliography}{99}
\bibitem{teleport} C. H. Bennett, G. Brassard, C. Cr\'epeau, R. Jozsa, A. Peres, and W. K. Wootters, Phys. Rev. Lett. \textbf{70}, 1895 (1993).
\bibitem{mbqc} R. Raussendorf and H.J. Briegel, Phys. Rev. Lett. \textbf{86}, 5188 (2001).
\bibitem{faulttolmbqc} M. A. Nielsen and C. M. Dawson, Phys. Rev. A \textbf{71}, 042323 (2005).
\bibitem{distillME}C. H. Bennett, G. Brassard, S. Popescu, B. Schumacher, J. A. Smolin, and W. K. Wootters, Phys. Rev. Lett. \textbf{76}, 722 (1996), Phys. Rev. Lett. \textbf{78}, 2031 (E) (1997).
\bibitem{distillg}W. D\"ur, H. Aschauer, and H.J. Briegel, Phys. Rev. Lett. \textbf{91}, 107903 (2003); H. Aschauer, W. D\"ur, and H.J. Briegel, Phys. Rev. A  \textbf{71}, 012319 (2005).
\bibitem{distillg2} C. Kruszynska, A. Miyake, H.J. Briegel, and W. D\"ur, Phys. Rev. A \textbf{74}, 052316 (2006).
\bibitem{faulttolmbqc2} R. Raussendorf, J. Harrington,  and K. Goyal, Ann. Phys. \textbf{321}, 2242 (2006).
\bibitem{faulttolmbqc3} C. M. Dawson, H. L. Haselgrove, and M. A. Nielsen, Phys. Rev. A \textbf{73}, 052306 (2006).
\bibitem{renyi} A. Re\'nyi, Proceedings of the Fourth Berkeley Symposium on
Mathematical Statistics and Probability, Volume 1: Contributions to the Theory of Statistics, 547 (1961).
\bibitem{resourceth} M. Horodecki, P. Horodecki, and J. Oppenheim, Phys. Rev. A \textbf{67}, 062104  (2003).
\bibitem{renyi2} G. Gour, M. P. M\"uller, V. Narasimhachar, R. W. Spekkens, and N. Y. Halpern, Phys. Rep. \textbf{583}, 1 (2015).
\bibitem{resoureth} A. Streltsov, H. Kampermann, S. W\"olk, M. Gessner, and  D. Bru\ss, New J. Phys. \textbf{20}, 053058 (2018).
\bibitem{distillpure} M. Horodecki, K. Horodecki, P. Horodecki, R. Horodecki, J. Oppenheim, A. Sen (De), U. Sen, Phys. Rev. Lett. \textbf{90}, 100402 (2003).
\bibitem{distillpure2} I. Devetak, Phys. Rev. A \textbf{71}, 062303 (2005).
\bibitem{finitevolume1}K. \.{Z}yczkowski, P. Horodecki, A. Sanpera, and M. Lewenstein, Phys. Rev. A \textbf{58}, 883 (1998).
\bibitem{finitevolume2} L. Gurvits and H. Barnum, Phys. Rev. A \textbf{66}, 062311 (2002).
\bibitem{enmeSC} M. A. Nielsen, Phys. Rev. Lett \textbf{83}, 436 (1999).
\bibitem{optimal1} A. Ac\'\i n, R. Tarrach, and G. Vidal, Phys. Rev. A \textbf{61}, 062307 (2000).
\bibitem{optimal2} E. Bagan, M. A. Ballester, R. Mu\~noz-Tapia and O. Romero-Isart, Phys. Rev. Lett. \textbf{95}, 110504 (2005).
\bibitem{Concboundl}F. Mintert and A. Buchleitner, Phys. Rev. Lett. \textbf{98}, 140505 (2007).
\bibitem{Concbound} C.-J. Zhang, Y.-X. Gong, Y.-S. Zhang, and G.-C. Guo, Phys. Rev. A \textbf{78}, 042308 (2008).
\bibitem{conc} S. Hill and W.K. Wootters, Phys. Rev. Lett. \textbf{78}, 5022 (1997).
\bibitem{conc2} P. Rungta, V. Bu\u{z}ek, C. M. Caves, M. Hillery, and G. J. Milburn
Phys. Rev. A \textbf{64}, 042315 (2001).
\bibitem{horodeck} R. Horodecki and M. Horodecki, Phys. Rev. A \textbf{54}, 1838 (1996).
\bibitem{exppur} A. K. Ekert, C. M. Alves, D. K. L. Oi, M. Horodecki, P. Horodecki, and L. C. Kwek, Phys. Rev. Lett. \textbf{88}, 217901 (2002).
\bibitem{exppur3} C. Moura Alves, and D. Jaksch, Phys. Rev. Lett. \textbf{93}, 110501 (2004).
\bibitem{exppur5} A. J. Daley, H. Pichler, J. Schachenmayer, and P. Zoller
Phys. Rev. Lett. \textbf{109}, 020505 (2012).
\bibitem{exppur2} H. Pichler, L. Bonnes, A. J. Daley, A. M. L\"auchli, and P. Zoller, New J. Phys. \textbf{15}, 063003 (2013).
\bibitem{exppur4} R. Islam, R. Ma, P. M. Preiss, M. E. Tai, A. Lukin, M. Rispoli,
and M. Greiner, Nature \textbf{528}, 77 (2015).
\bibitem{randmeas1} S. J. van Enk and C. W. J. Beenakker, Phys. Rev. Lett. \textbf{108}, 110503 (2012).
\bibitem{randmeas2} A. Elben, B. Vermersch, M. Dalmonte, J. I. Cirac, and P. Zoller, Phys. Rev. Lett. \textbf{120}, 050406 (2018).
\bibitem{randmeas3} A. Elben, B. Vermersch, C. F. Roos, and P. Zoller, Phys.
Rev. A \textbf{99}, 052323 (2019).
\bibitem{randmeas4} T. Brydges, A. Elben, P. Jurcevic, B. Vermersch, C. Maier, B. P. Lanyon, P. Zoller, R. Blatt, and C. F. Roos, Science \textbf{364}, 260 (2019).
\bibitem{odddim} T. Tanaka, G. Kimura, and H. Nakazato, Phys. Rev. A \textbf{87}, 012303 (2013).
\bibitem{discrimination1} C. Zhang, G. Wang, and M. Ying, Phys. Rev. A \textbf{75}, 062306  (2007).
\bibitem{discrimination2} S. Mal, T. Pramanik, and A. S. Majumdar. Phys. Rev. A \textbf{87}, 012105 (2013).
\bibitem{Gaussian} M. G. A. Paris, F. Illuminati, A. Serafini, and S. De Siena, Phys. Rev. A \textbf{68}, 012314  (2003). 
\bibitem{Gaussian2} G. Adesso, A. Serafini, and F. Illuminati,  Open Syst. Inf. Dyn. \textbf{12}, 189 (2005). 
\bibitem{adaptive} D. Ahn, Y. S. Teo, H. Jeong, F. Bouchard, F. Hufnagel, E. Karimi, D. Koutn\'{y}, J. \v{R}eh\'a\v{c}ek, Z. Hradil, G. Leuchs, and L. L. S\'{a}nchez-Soto, Phys. Rev. Lett. \textbf{122}, 100404 (2019).
\bibitem{adaptive2} X.-D. Yu and O. G\"uhne, Phys. Rev. A \textbf{99}, 062310 (2019).
\bibitem{adaptivepure1} D. Goyeneche, G. Ca\~nas, S. Etcheverry, E. S. G\'{o}mez,
G. B. Xavier, G. Lima, and A. Delgado, Phys. Rev. Lett. \textbf{115}, 090401 (2015).
\bibitem{adaptivepure2} Q. P. Stefano, L. Reb\'on, S. Ledesma, and C. Iemmi, arXiv:1903.05709 [quant-ph].
\bibitem{adaptive3} A. Zhang, Y. Zhang, F. Xu, L. Li and  L. Zhang, J. Phys. A: Math. Theor. \textbf{51} 395304 (2018).
\bibitem{boundvioBI} F. Verstraete and M. M. Wolf, Phys. Rev. Lett. \textbf{89}, 170401 (2002).
\bibitem{boundvioBI2} E. Santos, Phys. Rev. A \textbf{69}, 022305  (2004); Phys. Rev. A \textbf{70}, 059901(E) (2004). 
\bibitem{witboundMP} R. Chaves, J. Bohr Brask, and N. Brunner,  Phys. Rev. Lett. \textbf{115}, 110501 (2015).
\bibitem{seclength1} H. Aschauer, J. Calsamiglia, M. Hein, and H. J. Briegel, Quant. Inf. Comp. \textbf{4}, 383 (2004).
\bibitem{seclength2} 
N. Wyderka, and O. G\"uhne, arXiv:1905.06928 [quant-ph].
\bibitem{seclength3} 
C. Eltschka, and J. Siewert, arXiv:1908.04220 [quant-ph].
\bibitem{accessinfo} 
M. Dall'Arno, and F. Buscemi, IEEE Transactions on Information Theory \textbf{65}, 2614 (2019).
\bibitem{instrument} T. Heinosaari  and M. Ziman, \textit{The Mathematical Language of Quantum Mechanics} (Cambrigde University Press, 2011).
\bibitem{us} J. Hoffmann, C. Spee, O. G\"uhne, and C. Budroni, New J. Phys. \textbf{20}, 102001 (2018).
\bibitem{othertheories} C. Budroni, G. Fagundes, and M. Kleinmann, New J. Phys. \textbf{21}, 093018 (2019).
\bibitem{expus} C. Spee, H. Siebeneich, T. F. Gloger, P. Kaufmann, M. Johanning, M. Kleinmann, C. Wunderlich, and O. G\"uhne, New J. Phys. \textbf{22}, 023028 (2020).
\bibitem{concmulti1}A. R. R. Carvalho, F. Mintert, and A. Buchleitner, Phys. Rev. Lett. \textbf{93}, 230501 (2004).
\bibitem{concmulti2} L. Aolita and F. Mintert, Phys. Rev. Lett. \textbf{97}, 050501 (2006) .
\bibitem{Wootters} W. K. Wootters, Phys. Rev. Lett. \textbf{80}, 2245 (1998).
\end{thebibliography}
\end{document}